\documentclass[12pt,preprint]{aastex} \usepackage{epsfig}
\usepackage{amsmath, amssymb,bm} \shorttitle{ Accretion Outbursts In Circumplanetary Disks} \shortauthors{Lubow \& Martin}
\begin{document}

\title{ACCRETION OUTBURSTS IN CIRCUMPLANETARY DISKS}
\author{ S.~H. LUBOW AND  R.~G. MARTIN  }
\affil{Space Telescope Science Institute, Baltimore, MD 21218, USA }


\label{firstpage}

\begin{abstract}

 We describe a  model for the long term evolution of a circumplanetary disk
 that is fed mass from a circumstellar disk and contains regions of low
 turbulence (dead zones). We show that such disks can be subject
 to accretion driven outbursts, analogous to outbursts previously modeled 
 in the context of circumstellar disks to explain FU Ori phenomena.
Circumplanetary disks around a proto-Jupiter can undergo outbursts for infall accretion rates onto the disks in the range $\dot
{M}_{\rm infall} \approx 10^{-9}$ to $10^{-7}\,\rm M_\odot\,yr^{-1}$, typical
of accretion rates in the T Tauri phase. 
During outbursts, the accretion rate and disk luminosity increases by several orders of magnitude.
Most of the planet mass growth during planetary gas accretion may occur via 
disk outbursts involving gas that is considerably hotter than predicted by 
steady state models.  
For low infall accretion rates $\dot{M}_{\rm infall} \la 10^{-10} \, \rm M_\odot\,yr^{-1}$ that occur in late stages of disk accretion,
disk outbursts are unlikely to occur, even if dead zones are present.
Such conditions are favorable for the formation of icy satellites.   
\end{abstract}


\section{Introduction}

According to recent models, after a planet forms in a circumstellar
disk and its mass reaches that of the order of Neptune's, tidal forces
from the planet open a gap in the disk
\citep{lin84,lin86,dangelo02,bate03}. Material continues to flow from
the circumstellar disk through the gap and onto the planet
\citep{artymowicz96, kley99,lubow99,dangelo03, lubow06, ayliffe09}. The planetary radius
is much smaller than its Hill radius and, since gas flows into the
Hill sphere with some angular momentum, a circumplanetary disk forms. If modeled as a standard viscous
accretion disk, the circumplanetary disk is tidally truncated by the
star at a radius of about $0.4$ times the Hill radius
\citep{martin11a}.  Nearly all studies of the global structure of
circumplanetary disks have assumed them to be fully turbulent. In this {\it Letter}, we explore the long-term
consequences of dead zones in circumplanetary disks, regions that have
no turbulence near the midplane. In particular, we show that these
disks can undergo outbursts.

The structure of a circumplanetary disk depends upon the rate of 
angular momentum transport across the disk.
 The angular momentum transport allows
mass to move inwards and be accreted onto the central planet.
Angular momentum transport is driven by turbulence, of which there are
two main sources.  First, the magneto rotational instability (MRI)
drives turbulence if the gas is sufficiently well ionized to be
coupled to a magnetic field \citep[][]{balbus91}. In the hot inner
parts of the disk, the field is well coupled due to thermal ionization
and the MRI drives turbulence. However, further out where the disk is
cooler, the low ionization fraction causes magnetic field
instabilities to be suppressed \citep{gammie96,gammie98,turner08}. The
disk is said to form a dead zone at the midplane because it is not
turbulent in this region. However, the disk becomes layered because external radiation such as
cosmic rays or X-rays can ionize the surface layers, allowing them to
be turbulent \citep[e.g.,][]{ glassgold04}. The second type of turbulence is
driven by gravitational instability if the disk acquires enough mass
\citep{paczynski78,lodato04}.

Young circumstellar disks are thought to undergo outbursts (FU Ori phenomena) as
a consequence  of a sudden change in the level of turbulence
\citep{gammie96, armitage01, zhu09, martin11b}.
Because a dead zone inhibits inward flow onto the star, mass should
build up within the circumstellar disk. With sufficient mass, the
dead zone becomes gravitationally unstable. The instability causes
turbulent heating and  increased ionization of the gas, which can
trigger the MRI. This triggering causes a sudden increase in the disk turbulence,
resulting in an accretion outburst. After the outburst, the remaining
disk cools sufficiently for a dead zone to form again and the outburst
cycle repeats periodically. This process,
the gravo-magneto (GM) instability, can be understood as a limit cycle
\citep{martin11b}. The outbursts can be explained as transitions
between steady state disk solutions in a diagram that plots 
the disk effective temperature versus the total surface density at a
fixed radius. This graph is analogous to the well-known S-curve used to
explain dwarf-nova outbursts. There is a range of accretion rates for
which no steady disk solution exists and for these rates the disk is
GM unstable. 

Similar outbursts could also occur in circumplanetary disks.
\cite{martin11a} suggested that dead zone formation is even more
favorable in circumplanetary disks than in circumstellar disks
because the surface densities may be higher, while the temperatures remain
low enough to avoid thermal ionization.  In this work, for the first time we model the global
structure of a layered circumplanetary disk over long timescales for a
variety of parameters. In Section~\ref{runs} we describe the disk
model and the disk evolution.  In Section~\ref{obs} we discuss the
observability of a circumplanetary disk. Finally, in Section~\ref{concs} we discuss implications of the
layered circumplanetary disk model.

\section{Circumplanetary Disk Models}
\label{runs}

\subsection{Model Description}
\label{mod}

We apply the layered disk model initially described in \cite{armitage01} 
and further developed by \cite{zhu09} and \cite{martin11b}.  
Material falls onto the disk at a constant
rate $\dot{M}_{\rm infall}$ and is accreted onto the central planet through
the inner disk boundary at
a rate $\dot{M}$. 
The model accounts for the effects of disk turbulence by a viscosity. The disk vertical structure
consists of layers: outer magnetically turbulent surface layers on the upper and lower disk surfaces
and a midplane layer that can be either nonturbulent (dead) or turbulent (active) due to
the effects of disk magnetic
fields and/or self-gravity.  The disk temperatures in these layers are accounted for by a 
$T-\tau$ relation that is based on dust opacity. Disk heating is taken to be only
due to accretion.  External heating by the star or planet is ignored.  Heating by the star
may be ignored for planet orbital radii $> 1 \,\rm AU $, since its  contribution to the disk temperature is much less than that required
for thermal ionization $\sim 800 \, \rm K$.
 The disk
is taken to be in Keplerian orbit about the planet.
The model
does not account for the effects of  pressure or self-gravity
on the disk rotation.  
The planet is assumed to accept whatever gas is accreted
through the disk inner boundary without significant expansion into 
the disk, as in the run-away phase of gas accretion of the core accretion model \citep{dangelo10}.
We take the mass of the planet to be fixed and do not
account for its mass increase through gas accretion. 

 The disk has a total surface density $\Sigma(R,t)$,
midplane temperature $T_{\rm c}(R,t)$ and surface temperature $T_{\rm
  e}(R,t)$ for radius $R$ and time $t$. The MRI turbulent surface
layers have total surface density
$\Sigma_{\rm m}(R,t)$, temperature $T_{\rm m}(R,t)$, and viscosity
$\nu_{\rm m}(R,t)$ that is parameterized with the
\cite{shakura73} $\alpha_{\rm m}$ parameter. The external radiation is
 assumed to be able to penetrate to a constant surface
density $\Sigma_{\rm crit}$ in total for the disk upper and lower surface regions. If the surface density is greater
than this critical value, $\Sigma>\Sigma_{\rm crit}$, then
$\Sigma_{\rm m}=\Sigma_{\rm crit}$.  For smaller surface density,
$\Sigma<\Sigma_{\rm crit}$, the disk is MRI active at all heights, so that
$\Sigma_{\rm m}=\Sigma$.

A complementary midplane layer exists if the surface density is
larger than the critical value, $\Sigma>\Sigma_{\rm crit}$. It has surface
density $\Sigma_{\rm g}=\Sigma-\Sigma_{\rm crit}$ and midplane
temperature $T_{\rm c}$. This layer is MRI active if the midplane
temperature is greater than the critical, $T_{\rm c}>T_{\rm crit}$,
where $T_{\rm crit}$ is the temperature for sufficient thermal
ionization.  However, for lower temperatures, $T_{\rm c}<T_{\rm
  crit}$, a dead zone forms.  There is then no turbulence near the midplane
unless it is massive enough to be self-gravitating. The condition for
self-gravity is taken to be that the Toomre parameter \citep{toomre64} is smaller than
its critical value, $Q<Q_{\rm crit}=2$. In this case, a second
viscosity term is included in the complementary layer.

We consider the  case of a Jupiter-like planet of  mass, $M_{\rm p}= 1\, M_{\rm J}$,
orbiting a solar mass star, $M=1\,\rm M_\odot$, at a distance of
$a=5.2\,\rm AU$. We solve the accretion disk equations numerically on
a fixed mesh that is uniform in $R^\frac{1}{2}$ with 120 grid points
\citep[e.g.,][]{armitage01, martin07}. The inner boundary has a zero
torque condition at the radius of Jupiter, $R=1\,R_{\rm J}$, so the mass
falls freely onto the planet. At the outer boundary we approximate
the tidal torque from the star with a zero radial velocity boundary
condition at $R_{\rm out}=0.4\,R_{\rm H}$, equivalent to equation~70 of
\cite{martin11a}.
Circumstellar gas accretes onto the circumplanetary disk at a constant rate $\dot{M}_{\rm
  infall}$ at a radius $R_{\rm add}$.

The specific angular momentum of the accreting gas entering the planet's Hill
sphere from the circumstellar disk has yet to be determined. One estimate is obtained by
assuming that the gas enters near the Lagrange points, $ L_1$ and $L_2$,
and is nearly at rest relative to these points as it enters. By conservation of
angular momentum, the circularization radius for infalling material is
estimated to be about $0.33\,R_{\rm H}$ \citep{quillen98,estrada09}
and we take this radius to be $R_{\rm add}=0.33\,\rm R_{\rm H}$. We have also considered
cases of  $R_{\rm add}=0.1\,\rm R_{\rm H}$ and found that the results
are about the same. This outcome is consistent with the results
for a fully turbulent disk
by \cite{martin11a} that show the circumplanetary disk structure 
is not very sensitive to the angular momentum distribution
of the incoming circumstellar gas.
Initially we take the surface density to be very small everywhere and
 allow material to build up. We run the simulation until either the
disk reaches a steady state or the mass accretion onto the
central planet shows an outburst pattern that repeats itself.

The properties of the turbulent surface layer  have not been well
determined. There have been recent attempts to determine its depth, $\Sigma_{\rm crit}$,
by  calculating in detail the ionization balance of the disk due to external sources
of radiation and subject to various processes, such as ambipolar diffusion
and the presence of PAH and dust that tend to suppress the ionization \citep[e.g.,][]{bai09,p-b11,bai11}.
When applied to protostellar disks, these calculations result in accretion rates
that are lower than observed in typical T Tauri stars. The discrepancy
is currently unresolved.
Rather than attempt to fold in such an approach in this {\it Letter},
we follow Armitage et al. (2002) and Zhu et al. (2009, 2010) and regard
$\Sigma_{\rm crit}$ as a free parameter. Since typical T Tauri accretion
rates suggest that $\Sigma_{\rm crit} > 10\, \rm g\,cm^{-2}$ \citep{p-b11}, we consider values
in the range $100\, \rm g\,cm^{-2} \ge \Sigma_{\rm crit} \ge 10\, \rm g\,cm^{-2}.$

Following Armitage et al. (2001), we take $T_{\rm crit}=800\,\rm K$.
Various arguments suggest the viscosity $\alpha_{\rm m}$ parameter is found to be
$\sim 0.01-0.4$ \citep[e.g.][]{hartmann98,fromang07,guan09, davis09, zhu07, king07}. 
With the uncertainty in the value, we consider cases
with $\alpha_{\rm m}=0.01$ and $0.1$.

 During times of planet formation, the accretion rate onto the
 circumplanetary disk, $\dot{M}_{\rm infall}$, is of order the overall
 circumstellar disk accretion rate (Bate et al. 2003; Lubow \&
 D'Angelo 2006; Ayliffe \& Bate 2009) that we take to be of order
 $10^{-8}\,\rm M_\odot\,yr^{-1}$ \citep{valenti93,hartmann98}, typical
 for the T Tauri phase. In this {\it Letter}, we consider a range of values for
 $\alpha_{\rm m}$, $\dot{M}_{\rm infall}$, and $\Sigma_{\rm crit}$.

\subsection{Model Results \label{res}}

In Table~\ref{table} we show the outcome for a set of model
parameters. If the model does not reach a steady state and outbursts
occur, then we tabulate the time interval between outbursts, $t_{\rm int}$, and
the duration of outbursts, $t_{\rm out}$.
 For $\alpha_{\rm m}$ in the range $0.01-0.1$ with
$\Sigma_{\rm crit}\lesssim 100\,\rm g\,cm^{-2}$, outbursts are
possible for accretion rates onto the disk of $\dot{M}_{\rm
  infall}=10^{-9} $ to  $10^{-7}\,\rm M_\odot\,yr^{-1}$. They repeat on a typical
timescale of  $10^4$ to $10^5 \,\rm yr$ and last for a few to several years. 
We caution, however, that in the outburst state the disk thickness 
is formally of order the disk radius. Consequently, the idealization
of a disk geometry in Keplerian rotation at this stage is questionable. In the quiescent state, the disk thickness is 
a few tenths of the disk radius, as expected by equation 4 of \cite{martin11a} and the three-dimensional simulations
of \cite{ayliffe09}. We find that the disk mass reaches values that are not very small
compared to the planet mass. For example, model R3 reaches values as high as about 0.3 $M_{\rm J}$.
The Keplerian disk assumption is then not well satisfied at all times. Improved multi-dimensional modeling
should be carried out in future work.

The results indicate 
a maximum accretion rate, $\dot{M}_{\rm max}$, of about
$10^{-4}\,\rm M_\odot\,yr^{-1}$ and a quiescent accretion rate of
about $10^{-9}\,\rm M_\odot\, yr^{-1}$. Consequently,  accretional heating
of the gas increases substantially during an outburst. 
The mass accreted during an outburst is somewhat greater than
the mass accreted between outbursts, typically 30\% to 40\% greater.
The disk is stabilized against outbursts with a larger active outer layer surface density,
$\Sigma_{\rm crit}$, or stronger turbulence $\alpha_{\rm m}$. 
For smaller values of $\Sigma_{\rm crit}$, outbursts become more problematic.
Model R7 with  $\Sigma_{\rm crit}=10   \, \rm g\,cm^{-2}$ has a longer interval between outbursts
and involves an accreted mass during outburst  that is comparable to that of the planet which
suggests a more detailed treatment is required to determine whether outbursts occur in this regime.
The detailed ionization models discussed in Section~\ref{mod}  obtain even lower values of $\Sigma_{\rm crit}$,
for which our models are subject to this limitation.

\begin{figure}
\begin{center}
\includegraphics[width= 12 cm]{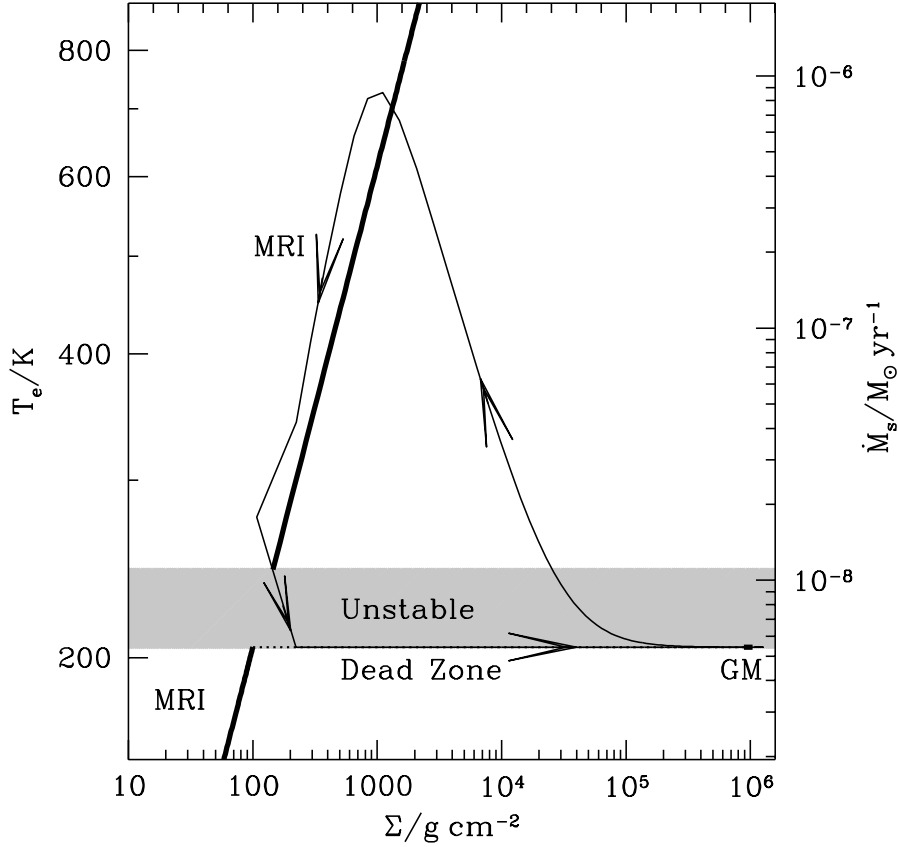}
\end{center}
\caption{Limit cycle for model R3 at a radius of $R=0.07\,R_{\rm
    H}$ in the $\Sigma-\dot{M}_{\rm s}$ state or $\Sigma-T_{\rm e}$  diagram. 
    $T_{\rm e}$ is the disk surface temperature that is less than $T_{\rm c}$, the
    temperature at the disk midplane that controls the onset of MRI via thermal ionization.
  The fully turbulent (no dead zone) solutions involving MRI  have two
  branches. They are plotted as the sloped heavy line with a gap. The lower branch (below the gap) is ionized throughout by external sources of radiation and the
  upper branch is thermally ionized. The steady gravo-magneto (GM)
  solution is shown as a short segment on the right hand end of the horizontal line. The dotted line shows
  where the nonsteady dead zone branch resides that connects the lower
  MRI branch to the GM branch. The thin line shows the simulated
  evolution in time that evolves in a counter-clockwise direction. It extends horizontally to the GM branch. The
  shaded region shows the range of accretion rates for which the disk
  has no steady solution and would trigger an outburst. }
\label{arm2}
\end{figure}

The instability cycle can be understood through a
state transition diagram that plots
effective surface temperature (or equivalent steady state accretion rate) as a function
of disk surface density at some fixed radius in the disk  \citep{martin11b}.
In Fig.~\ref{arm2} we show the limit cycle for model R3 in the
such a diagram at a radius
$R=0.07\,R_{\rm H}$. The steady accretion rate, $\dot{M}_{\rm s}$,
corresponds to the surface temperature of the disk given by
equation~18 in \cite{martin11b}. The thick solid lines show the two
types of analytical steady state solutions. The first is fully MRI
active and this has two branches, one with $T>T_{\rm crit}$ (upper
branch) and a second with $\Sigma<\Sigma_{\rm crit}$ (lower
branch). The second type of steady state is the gravo-magneto (GM)
solution that has a self gravitating dead zone and MRI active
surface layers.  The dotted line shows the nonsteady dead zone branch
that connects the lower MRI branch to the GM branch.  The numerical
simulations reveal a cycle on this $\Sigma-\dot{M}_{\rm s}$
diagram. The  simulation results plotted in Fig.~\ref{arm2} generally
track  along the dead zone and steady state branches. 
The simulation track moves to the right from the dead zone branch to  the GM branch. 
This branch is plotted as a short line
segment at the right end of the horizontal line. After outburst, the simulation
backtracks somewhat to the left and slightly upward, and then  advances directly upward to the upper MRI branch.
This upward transition marks the onset of the outburst.

Between the dead zone/GM branches and the upper MRI branch
there is a range of accretion rates $\dot{M}_{\rm s}$ for which
no steady state is possible and an outburst results.  
This unstable regime is plotted as the shaded region.
Model R3 lies in that range at the selected radius, since it has an infall accretion rate of $10^{-8} \, \rm M_\odot\,yr^{-1}$.
The global range of unstable accretion rates covers a somewhat wider range than is shown on the plot,
since other unstable rates may occur at other radii.


The results in Table~\ref{table} are consistent with there being a limited range
of infall accretion rates  that result in outbursts.
Model R10 undergoes outbursts while models R9 and R11, having
lower and higher infall accretion rates respectively, do not undergo outbursts,
with all other input parameters being the same.


\begin{table*}
\label{aggiungi}\centering
\begin{tabular}{cclcccccccccc}
\hline
Model   &  $\alpha_{\rm m}$ &  $\dot{M}_{\rm infall}$& $\Sigma_{\rm crit}$  & $t_{\rm int}/\rm yr$  &  $t_{\rm dur}/\rm yr$ & $M_{\rm burst}/\rm M_{\rm J}$  & $\dot{M}_{\rm min}/{\rm M_\odot\,yr^{-1}}$& $\dot{M}_{\rm max}/{\rm M_\odot\,yr^{-1}}$ \\ 
\hline
R1	&	0.01  &      $10^{-10}$	       &	40	&	 {\rm steady} \\
R2 	&	0.01  &      $10^{-9}$	       &	40	&	 $1.38\times 10^5$ & $5.81$ & $0.065$ & $3.63\times 10^{-10}$  & $9.46\times 10^{-5}$  \\ 
R3 	&	0.01  &      $10^{-8}$	       &	100	&	 $1.67\times 10^4$ &  $12.4$  &  $0.096$ & $4.60\times 10^{-9}$  & $4.12\times 10^{-5}$  \\
R4 	&	0.01  &      $10^{-8}$	       &	200	&	 {\rm steady} \\
R5 	&	0.01  &      $10^{-6}$	       &	40	&	 {\rm steady} \\
R6	&	0.1  &      $10^{-9}$	       &	20	&	 {\rm steady} \\
 R7   &	0.1  &      $10^{-8}$	       &	10	&	 $8.56\times 10^4$ & $5.39$ & $0.66$ & $6.58\times 10^{-10}$  & $1.8\times 10^{-3}$ \\ 
 R8 	&	0.1  &      $10^{-8}$	       &	20	&	 $1.66\times 10^4$ & $1.38$ & $0.087$ & $4.00\times 10^{-9}$  & $6.9\times 10^{-4}$ \\ 
 R9 	&	0.1  &      $10^{-8}$	       &	40	&	 {\rm steady} \\ 
  R10 &	0.1  &      $10^{-7}$	       &	40 & $3.48\times 10^3$ & 1.5 &  $0.143$ & $2.92\times 10^{-8}$  & $3.54\times 10^{-4}$  \\ 
R11	&	0.1  &      $10^{-6}$	       &	40	& {\rm steady} \\
\hline
\end{tabular}
\caption{The parameters of each model and the timescales of the
  outbursts if they occur. Column~2 is the $\alpha_{\rm m}$ parameter
  in the active outer layers, column~3 is the accretion rate onto the disk in units of ${\rm M_\odot\,yr^{-1}}$,
  column~4 is the total surface density of the  active outer layers (in $\rm
  g\,cm^{-2}$), 
  column~5 is the time interval between outbursts, column~6 is the duration
  of the outburst, column~7 is the mass accreted onto the planet
  during the outburst, column~8 is the accretion rate during
  quiesence and column~9 is the maximum accretion rate onto the
  planet achieved during the outburst.}
\label{table}
\end{table*}

\section{Observability of a Circumplanetary Disk}
\label{obs}

The average accretion rate onto the planet is equal to the rate of
infall accretion onto the disk.  Because the accretion through the
disk is time dependent, so is the luminosity. The luminosity of an
accretion disk is
\begin{equation}
L=\frac{G M \dot{ M}}{2 R_{\rm in}}
\end{equation}
\citep{pringle81}, where $R_{\rm in}$ is the inner edge of the disk
that is assumed to coincide with the radius of the central star or
planet. For example, a disk around a solar mass and radius star with
an accretion rate of $10^{-8}\,\rm M_\odot\,yr^{-1}$ has an accretion luminosity
of $0.15\,\rm L_\odot$ (in addition to the luminosity due to heating from the star). 
 If a
disk surrounds a Jupiter mass and (constant) Jupiter radius planet that has opened a gap
in the circumstellar disk, the quiescent luminosity of the
circumplanetary disk is $0.0015\,\rm L_\odot$ or only 1 percent  of
accretion luminosity of the circumstellar disk. 

 If the disk is observed during an outburst
with a high accretion rate of $10^{-4}\,\rm M_\odot\, yr^{-1}$, then
the luminosity is $\sim 15\,\rm L_\odot$ (see Table~\ref{table}). However,  the disk will be in outburst only
for $0.01 \%-0.1 \%$ of time. It is possible that with the increased luminosity even
for short periods of time, forming planets may be more easily detected.
Synoptic surveys may be capable of detecting such events in the future.

\section{Discussion and Conclusions}
\label{concs}

Circumplanetary disks can contain regions
with low turbulence (dead zones) because they have high enough surface densities
to prevent ionization of the disk midplane by external sources
of radiation, but remain cool
enough to avoid thermal ionization. 
As a result of dead zone formation, the disks are unstable to the
gravo-magneto (GM) instability in which a sudden turn-on of the MRI
triggers a rapid accretion event (see Fig.~1). 
We have found that such outbursts are possible
for infall accretion rates onto the disk in
the range of $10^{-9}$ to $10^{-7}\,\rm M_\odot\,\rm yr^{-1}$, typical of
accretion rates in the T Tauri phase of young stars, when gas giant planet
formation may occur.
 The quiescent accretion rates onto  planets are $ \sim 10^{-9}\,\rm
M_\odot\,yr^{-1}$, but during outbursts the rates increase up to 
$\sim 10^{-4}\,\rm M_\odot\,yr^{-1}$, making them more observable. 
The outbursts typically occur every $\sim 10^4$ to $10^5 \, \rm{yr}$ and
have a duration of only several years. 
Somewhat more gas is accreted during outbursts
than between them. 

 The results have implications
for the nature of gas accretion onto a planet during
the run-away gas
accretion phase of the core accretion model (see D'Angelo et al. 2010).
 The higher entropy we expect to be associated with the gas accreted onto a planet in
 outbursts may have implications on the initial state of a newly
 formed planet. The higher entropy  may favor more luminous "hot start" planet models (c.f. Marley et al 2007). 

There are several approximations in the layered disk model we have
considered that should be improved upon in future work. 
Some of these
were discussed in \cite{martin11b} in the context of circumstellar disks. 
The present treatment is based on a one-dimensional model of a disk in Keplerian rotation.
As noted in Section \ref{res}, in the case of circumplanetary disks, the disk thicknesses and masses
can result in  
structures that may not be well approximated by a Keplerian
disk. Multi-dimensional simulations that include pressure and self-gravity would be of benefit
in understanding
the evolution of the outburst and the accretion
onto the planet.

Lower accretion rates ($\la 10^{-10}  {\rm M_\odot\,yr^{-1}}$) provide lower disk temperatures that
would be compatible with the survival of icy satellites at their current locations \citep{canup02,mosqueira03,ward10}.
Such rates can occur at late stages of disk evolution. 
The Jovian circumplanetary disk could have contained a dead zone in the regular satellite region,
provided that external sources of radiation penetrated a disk surface density of $\Sigma_{\rm crit} \la 10\,\rm g\,cm^{-2}$.
It is unlikely that outbursts will occur at this stage even in the presence of a dead zone, since the time required for a disk to reach a self-gravitating
state is longer than the disk lifetime.
Therefore, although outbursts could adversely affect the survival of icy satellites, they are unlikely to occur
at these late stages.

\section*{Acknowledgements}
  SHL acknowledges support from NASA grant NNX07AI72G.
RGM thanks the Space Telescope Science Institute for a Giacconi
Fellowship.


\label{lastpage}

\begin{thebibliography}{99}

\bibitem[Armitage at al. (2001)]{armitage01}
Armitage P. J., Livio M., Pringle J. E., 2001, MNRAS, 324, 705

\bibitem[Artymowicz \& Lubow(1996)]{artymowicz96}
Artymowicz P., Lubow S. H., 1996, ApJ, 467, L77 

\bibitem[Ayliffe \& Bate(2009)]{ayliffe09}
Ayliffe B. A., Bate M. R., 2009, MNRAS, 393, 49

\bibitem[Balbus \& Hawley(1991)]{balbus91}
Balbus S. A., Hawley J. F., 1991, ApJ, 376, 214

\bibitem[Bai (2011)]{bai11}
Bai, X.-N. 2011, ApJ, 739, 50

\bibitem[Bai \& Goodman (2009)]{bai09}
Bai, X.-N., Goodman, J. 2009, ApJ, 701, 737

\bibitem[Bate et al.(2003)]{bate03}
Bate M. R., Lubow S. H., Ogilvie G. I., Miller K. A., 2003, MNRAS, 341, 213 



\bibitem[Canup \& Ward(2002)]{canup02}
Canup R. M., Ward W. R., 2002, ApJ, 124, 3404 



\bibitem[D'Angelo, Henning \& Kley(2002)]{dangelo02}
D'Angelo G., Henning T., Kley W., 2002, A\&A, 385, 647 

\bibitem[D'Angelo, Henning \& Kley(2003)]{dangelo03}	
D'Angelo G., Henning T., Kley W., 2003, ApJ, 599, 548

\bibitem[D'Angelo, Durisen, \& Lissauer (2010)]{dangelo10} D'Angelo, G., Durisen, 
R.~H., \& Lissauer, J.~J.\ 2010, Exoplanets, 319 


\bibitem[Davis et al.(2009)]{davis09}
Davis S. W., Blaes O. M., Hirose S., Krolik J. H., 2009, ApJ, 703, 569 

\bibitem[Estrada et al.(2009)]{estrada09} Estrada P. R., Mosqueira
  I. L., Lissauer J. J., D'Angelo G., Cruikshank D. P., 2009, in
  Europa, eds Pappalardo R. T., McKinnon W. B., Khurana K. K.,
  University of Arizona Press, Tucson, 27


\bibitem[Fromang et al.(2007)]{fromang07}
Fromang S., Papaloizou J., Lesur G., Heinemann T., 2007, A\&A, 476, 
1123 


\bibitem[Gammie(1996)]{gammie96}
Gammie C. F., 1996, ApJ, 457, 355

\bibitem[Gammie \& Menou(1998)]{gammie98}
Gammie C. F., Menou K., 1998, ApJ, 492, 75

\bibitem[Glassgold, Najita \& Igea(2004)]{glassgold04}
Glassgold A. E., Najita J., Igea J., 2004, ApJ, 615, 972 

\bibitem[Guan et al.(2009)]{guan09}
Guan X., Gammie C. F., Simon J. B., Johnson B. M., 2009, ApJ, 694, 1010 

\bibitem[Hartmann et al.(1998)]{hartmann98}
Hartmann L., Calvet N., Gullbring E., D’Alessio P., 1998, ApJ, 495, 385 

\bibitem[King, Pringle \& Livio(2007)]{king07}
King A. R., Pringle J. E., Livio M., 2007, MNRAS, 376, 1740

\bibitem[Kley(1999)]{kley99}
Kley W., 1999, MNRAS, 303, 696


\bibitem[Lin \& Papaloizou(1984)]{lin84}
Lin D. N. C., Papaloizou J.,1984, ApJ, 285, 818

\bibitem[Lin \& Papaloizou(1986)]{lin86}
Lin D. N. C., Papaloizou J.,1986, ApJ, 309, 846 



\bibitem[Lodato \& Rice(2004)]{lodato04}
Lodato G., Rice W. K. M., 2004, MNRAS, 351, 630

	

\bibitem[Lubow, Seibert \& Artymowicz(1999)]{lubow99}
Lubow S. H., Seibert M., Artymowicz P., 1999, ApJ, 526, 1001 

\bibitem[Lubow \& D'Angelo(2006)]{lubow06}
Lubow S. H., D'Angelo G., 2006, ApJ, 641, 526


\bibitem[Marley et al.(2007)]{2007ApJ...655..541M} Marley, M.~S., Fortney, 
J.~J., Hubickyj, O., Bodenheimer, P., 
\& Lissauer, J.~J.\ 2007, \apj, 655, 541 


\bibitem[Martin et al.(2007)]{martin07} Martin R. G., Lubow S. H.,
  Pringle J. E., Wyatt M. C., 2007, MNRAS, 378, 1589

\bibitem[Martin \& Lubow(2011a)]{martin11a}
Martin R. G., Lubow S. H., 2011a, MNRAS, 413, 1447

\bibitem[Martin \& Lubow(2011b)]{martin11b} Martin R. G., Lubow S. H.,
  2011b, ApJ, 740, L6





\bibitem[Mosqueira \& Estrada (2003)]{mosqueira03}
 Mosqueira
  I. L.,Estrada P. R.,  2003, Icarus, 163, 198



\bibitem[Paczynski(1978)]{paczynski78}
Paczynski B., 1978, AcA, 28, 91

\bibitem[Perez-Becker \& Chiang (2011)]{p-b11}
Perez-Becker, D., Chiang, E. 2011, ApJ, 727, 2


\bibitem[Pringle(1981)]{pringle81}
Pringle J. E., 1981, ARA\&A, 19, 137

\bibitem[Quillen(1998)]{quillen98}
Quillen A. C., Trilling D. E., 1998, ApJ, 508, 707 

\bibitem[Shakura \& Sunyaev(1973)]{shakura73}
Shakura N. I., Sunyaev R. A., 1973, A\&A, 24, 337


\bibitem[Toomre (1964)]{toomre64}
Toomre, A. 1964, ApJ, 139, 1217





\bibitem[Turner \& Sano(2008)]{turner08}
Turner N. J., Sano T.,  2008, ApJ, 679, 131


\bibitem[Valenti, Basri \& Johns(1993)]{valenti93}
Valenti J. A., Basri G., Johns C. M., 1993, AJ, 106, 2024


\bibitem[Ward \& Canup(2010)]{ward10}
Ward W. R., Canup R.M., 2010, AJ, 140, 1168 


\bibitem[Zhu et al.(2007)]{zhu07}
Zhu Z., Hartmann L., Calvet, N., Hernandez J., Muzerolle J., Tannirkulam 
A. K. 2007, ApJ, 669, 483 

\bibitem[Zhu et al.(2009)]{zhu09}
Zhu Z., Hartmann L., Gammie C., 2009, ApJ, 694, 1045

\bibitem[Zhu et al.(2010)]{zhu10} Zhu Z., Hartmann L., Gammie C. F.,
  Book L. G., Simon J. B., Engelhard E., 2010, ApJ, 713, 1142

\end{thebibliography}
\end{document}